# Claim Plane: Reliability Gains and the Limits of Selective Concurrency for Parallel Coding Agents

## A 30-Pair, Three-Seed Confirmatory Study of Deterministic Pre-Write Admission

**Maxim Nikolaev**
Vladivostok State University, Vladivostok, Russia
Email: m.nikolaev.contact@gmail.com | ORCID: 0009-0009-5804-8270

**Abstract**

Parallel coding agents can produce locally valid changes that fail when combined. Claim Plane addresses this failure mode as deterministic pre-write admission over versioned change intents. This paper reports a confirmatory study on 30 frozen CooperBench feature pairs, balanced between 15 conflict and 15 clean labels, with three coder seeds, four coordination arms, and 360 completed executions. DeepSeek V4 Pro generated 60 feature-level planner declarations once; the declarations were frozen across all arms and coder seeds, while DeepSeek V4 Flash performed the coding work. Static Claim Plane raised pair pass from 23.3% under unconstrained parallel execution to 50.0%, a paired task-cluster difference of +26.7 percentage points (95% bootstrap CI 9.6 to 60.0), and raised integration success from 65.6% to 96.7%. On conflict-labeled pairs, pair pass rose from 6.7% to 60.0%. However, static admission serialized 96.7% of executions, including 93.3% of clean cases, and therefore recovered reliability largely by collapsing toward serial execution. Dynamic admission was more selective, serializing 66.7% of conflict cases and 13.3% of clean cases, but 46 of 90 executions failed closed on undeclared scope, reducing pair pass to 22.2%. Forty-five of those 46 blocks targeted files already present in the frozen declarations, indicating region undercoverage and insufficient amendment handling rather than wholly unknown files. The results support pre-write admission as a reliability mechanism, but they do not establish useful wall-clock parallel speedup: provider calls were physically sequential, and the conservative policy largely serialized the workload. The complete study artifacts, hashes, and clustered bootstrap analysis are publicly released for reproduction.

**Keywords:** parallel coding agents; claim-based admission; pre-write coordination; dynamic scope; CooperBench; verification; selective concurrency

## 1. Introduction

Repository-scale coding agents now edit multiple files, run test suites, and construct changes that span interfaces and internal contracts. When several agents work on the same repository, the systems problem is no longer only whether each model can implement its assigned feature. It is whether independently plausible changes remain mutually valid when executed and integrated together.

CooperBench makes this coordination penalty explicit by assigning two agents distinct features that may be individually solvable but jointly incompatible [2]. The benchmark reports substantial degradation when agents work together. Recent systems respond through dependency-aware delegation and isolated workspaces [3], cohesion-aware task partitioning [4], reversible execution traces and live supervision [5], shared-state convergence [6], or optimistic notification and repair [7]. These approaches show that coordination belongs in the execution architecture rather than only in natural-language conversation.

Claim Plane occupies a narrower point in this design space. It treats concurrent software change as a pre-write authority problem [1]. Before implementation, each worker declares a versioned ChangeIntent containing an exact base revision, typed resources, dependencies, and operations marked committed or contingent. A deterministic control plane admits compatible intent combinations, serializes or rejects unresolved overlap, and requires scope expansion to be re-admitted before the associated mutation proceeds.

The first Claim Plane paper presented the architecture and a six-pair mechanism check that was explicitly too small for comparative claims [1]. The present work is the pre-specified confirmatory follow-up. It freezes the planner output once, varies only three coding-agent seeds, compares four arms across 30 balanced feature pairs, and aggregates all 360 executions with task-cluster bootstrap intervals and a hashed publication manifest.

The paper makes four contributions:

- A 360-execution confirmatory evaluation of claim-based admission on 30 frozen CooperBench pairs.
- Evidence that conservative pre-write admission can recover always-serial reliability on conflict-labeled work.
- A failure analysis showing that selective dynamic admission is limited primarily by region-level scope undercoverage and amendment handling.
- A complete reproducibility record containing frozen plans, nine resumable shards, result rows, clustered bootstrap intervals, cost summaries, and hashes.

## 2. Claim Plane and Evaluated Policies

### 2.1 Enforceable change intent

Claim Plane separates probabilistic planning from deterministic authority. A planner, human, or external runtime may propose a ChangeIntent, but the control plane owns admission and mutation authority. Committed scope participates in admission immediately. Contingent scope expresses plausible future work without reserving write authority until a concrete mutation requires promotion. The full protocol, broker boundary, leases, fencing tokens, dependency invalidation, and immutable evidence chain are described in the companion design paper [1].

### 2.2 Static and dynamic policies

The static policy treats the calibrated final declaration as committed before coding. If two declarations contain unresolved overlap, the pair is serialized. This maximizes scope recall but can serialize clean work when declarations are broad or imprecise.

The dynamic policy initially reserves only committed operations. A mutation covered by contingent scope triggers atomic promotion and re-admission; a mutation outside both committed and contingent scope is blocked. This can preserve concurrency, but it depends on planner coverage and on an operational amendment path that can distinguish legitimate scope growth from unauthorized expansion.

### 2.3 Research questions

The confirmatory study addresses four research questions:

- RQ1 - Reliability: Does claim-based admission improve pair pass and integration success relative to unconstrained parallel execution?
- RQ2 - Selectivity: Can a policy serialize conflict-labeled pairs while preserving concurrency on clean-labeled pairs?
- RQ3 - Failure mechanism: Which integration, feature-test, execution, and scope-enforcement failures explain the observed outcomes?
- RQ4 - Economics and latency: What logical model cost and critical-path trade-offs accompany each coordination policy?

## 3. Experimental Method

### 3.1 Frozen study population

The study contains 30 feature pairs from seven CooperBench tasks in three Python project families: Jinja, Click, and dirty-equals. Pair selection used seed 42 and was balanced between 15 benchmark conflict labels and 15 clean labels. All 30 selected pairs passed the benchmark gold-sanity procedure for each individual feature before model execution. The CooperBench source revision was d46d9e73fa64159e0428b480f293623de90be1ad, and the frozen dataset SHA-256 was c5a7638846b4079a4e31cc2c40d7fad2728194c8b165c8f553211f39a7850cac.

### 3.2 Models, seeds, and frozen planning

Planner v1 used deepseek/deepseek-v4-pro. It generated 60 feature-level declarations once under planner freeze seed 1701. Those plans were reused across coder seeds 101, 202, and 303 and across the static and dynamic Claim Plane arms. This design prevents planner variation from being confounded with policy effects. The one-time frozen-planner logical cost was $1.9971.

Coding executions used deepseek/deepseek-v4-flash. Each of the 30 pairs was evaluated under four arms for each of three seeds, producing 30 x 4 x 3 = 360 arm executions. The experiment was divided into three ten-pair shards per seed, resulting in nine resumable run directories.

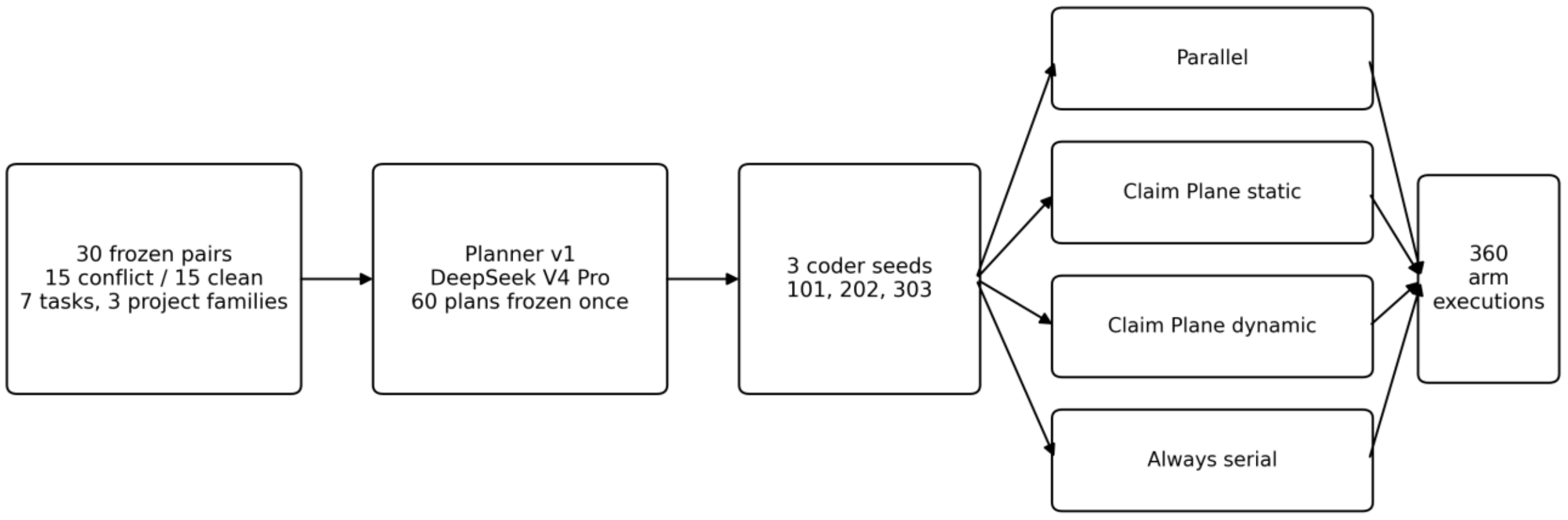

***Figure 1. Confirmatory study design. Planner declarations were frozen once and reused across three coder seeds and four coordination arms.***

### 3.3 Execution arms

| Arm | Execution policy | Role |
|---|---|---|
| Parallel | No Claim Plane admission; both feature workers are logically concurrent. | Unconstrained baseline |
| Claim Plane static | All calibrated declaration scope is committed before execution; unresolved overlap serializes the pair. | Conservative admission |
| Claim Plane dynamic | Committed scope is admitted initially; contingent mutations require promotion; undeclared writes fail closed. | Selective admission |
| Always serial | Feature workers execute in a fixed sequential order. | Reliability reference |

***Table 1. Coordination arms used in the confirmatory study.***

### 3.4 Outcomes and mechanism metrics

The primary outcome is pair pass: both individual feature evaluations pass and the integrated repository passes the combined benchmark checks. Integration success is reported separately because a branch pair can merge and integrate while still failing one or both feature requirements. Mechanism metrics include initial and effective serialization, promotion attempts, successful promotions, undeclared-scope blocks, wasted dynamic steps, and wasted coder cost. Cost reporting distinguishes coder cost from a logical system-cost estimate that allocates the frozen planner overhead to the Claim Plane deployment paths.

The benchmark supplied oracle-localized initial context derived from gold-patch locations. This condition reduces context-retrieval variance but limits external validity. API requests were physically sequential. Consequently, this paper does not claim measured provider-level wall-clock speedup; concurrency is evaluated through admission decisions and logical critical-path estimates.

### 3.5 Statistical analysis and reproducibility

The publication analysis uses a nonparametric percentile bootstrap with 5,000 samples, seed 20260727, and a 95% interval. The resampling unit is repository plus task identifier. This preserves dependence among related feature pairs and all coder seeds from the same task cluster. Only seven clusters are available, so intervals remain intentionally wide and are interpreted as uncertainty summaries rather than asymptotic guarantees.

The final publication manifest contains 360 arm executions, nine run identifiers, hashes for 20 analysis inputs and 15 outputs, and the study fingerprint. Verification completed with no mismatches.

## 4. Results

### 4.1 Main reliability outcomes

| Arm | Pair pass | Rate [95% CI] | Integration | Rate [95% CI] | Effective serialization |
|---|---|---|---|---|---|
| Parallel | 21/90 | 23.3% [3.3, 49.5] | 59/90 | 65.6% [31.6, 85.1] | 0/90 (0.0%) |
| Claim Plane static | 45/90 | 50.0% [25.0, 78.7] | 87/90 | 96.7% [90.3, 100.0] | 87/90 (96.7%) |

| Arm | Pair pass | Rate [95% CI] | Integration | Rate [95% CI] | Effective serialization |
|---|---|---|---|---|---|
| Claim Plane dynamic | 20/90 | 22.2% [7.1, 43.9] | 42/90 | 46.7% [23.8, 77.1] | 36/90 (40.0%) |
| Always serial | 45/90 | 50.0% [25.0, 78.7] | 86/90 | 95.6% [88.9, 100.0] | 90/90 (100.0%) |

***Table 2. Main outcomes. Confidence intervals use the repository-task cluster bootstrap.***

Static Claim Plane increased pair pass from 21/90 (23.3%) to 45/90 (50.0%). Because the static and always-serial arms had identical pair-pass rates within every repository-task cluster, the static-versus-parallel paired cluster difference equals +26.7 percentage points, with a 95% bootstrap interval from +9.6 to +60.0. Integration success increased by 31.1 points, from 65.6% to 96.7%. Static Claim Plane was never worse than parallel execution at the task-cluster level for the primary outcome and matched the always-serial aggregate pair-pass rate.

Dynamic Claim Plane did not improve end-to-end reliability. It achieved 20/90 pair passes (22.2%) and 42/90 successful integrations (46.7%). Its paired differences relative to always serial were -27.8 points for pair pass (95% CI -39.1 to -14.1) and -48.9 points for integration success (95% CI -71.9 to -20.4).

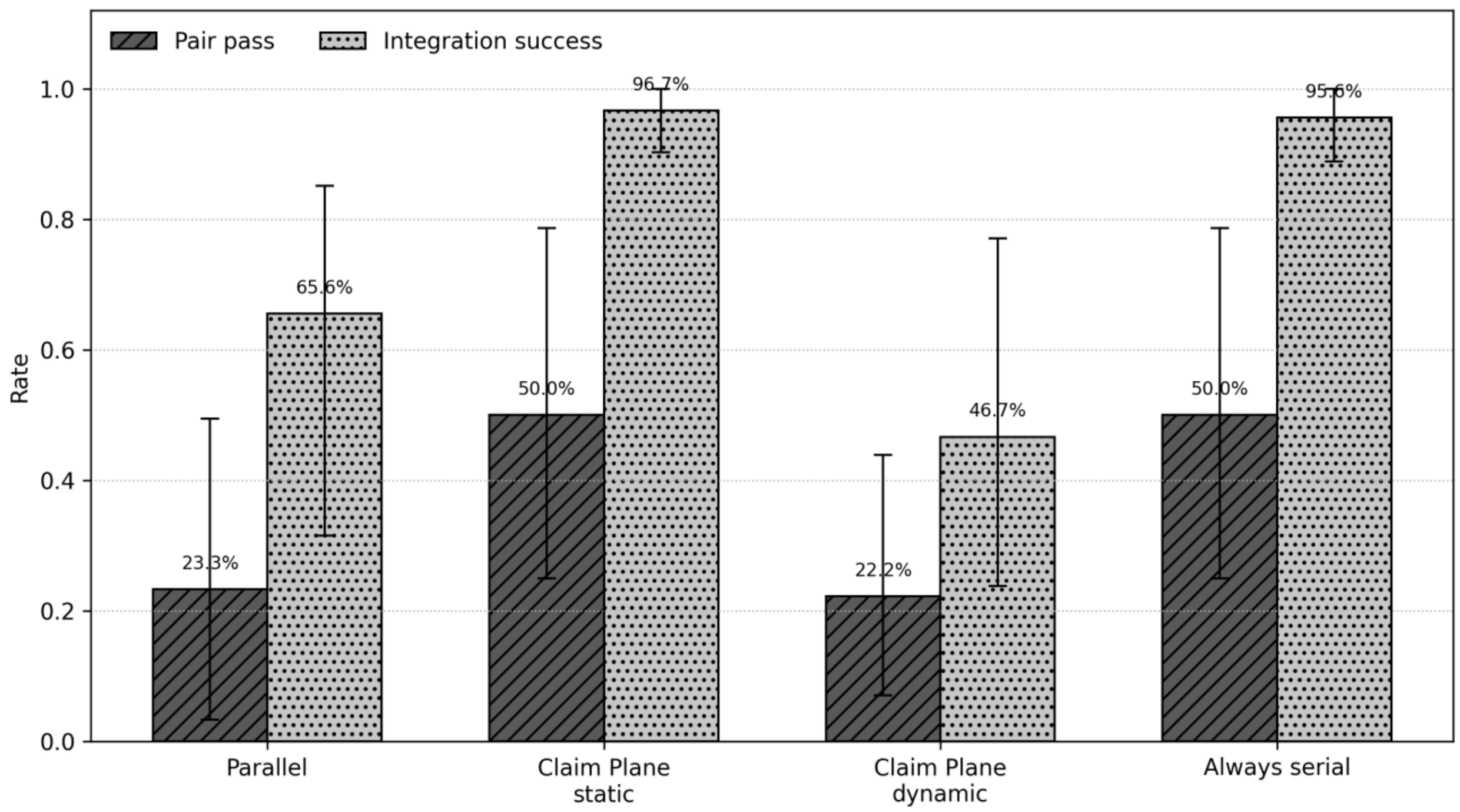

***Figure 2. Pair-pass and integration-success rates by arm. Error bars show 95% repository-task cluster bootstrap intervals.***

## 4.2 Conflict-labeled and clean-labeled strata

| Arm | Conflict pair pass | Conflict integration | Conflict serialized | Clean pair pass | Clean integration | Clean serialized |
|---|---|---|---|---|---|---|
| Parallel | 3/45 (6.7%) | 15/45 (33.3%) | 0/45 (0.0%) | 18/45 (40.0%) | 44/45 (97.8%) | 0/45 (0.0%) |
| Static | 27/45 (60.0%) | 43/45 (95.6%) | 45/45 (100.0%) | 18/45 (40.0%) | 44/45 (97.8%) | 42/45 (93.3%) |
| Dynamic | 15/45 (33.3%) | 23/45 (51.1%) | 30/45 (66.7%) | 5/45 (11.1%) | 19/45 (42.2%) | 6/45 (13.3%) |
| Always serial | 27/45 (60.0%) | 43/45 (95.6%) | 45/45 (100.0%) | 18/45 (40.0%) | 43/45 (95.6%) | 45/45 (100.0%) |

***Table 3. Outcomes stratified by CooperBench conflict and clean labels.***

The reliability gain was concentrated in conflict-labeled work. Parallel execution passed only 3/45 conflict pairs (6.7%) and integrated 15/45 (33.3%). Static Claim Plane passed 27/45 conflict pairs (60.0%) and integrated 43/45 (95.6%), exactly matching always serial in this stratum. On clean pairs, static and parallel execution had the same pair-pass rate (40.0%) and integration rate (97.8%). This supports a coordination interpretation: static Claim Plane prevented interference but did not improve the coding model itself.

## 4.3 Serialization selectivity

Static admission serialized every conflict-labeled execution and 42/45 clean executions. Its conflict sensitivity was therefore 100%, but clean specificity was only 6.7%. Dynamic admission was substantially more selective: it serialized 30/45 conflict executions (66.7%) and only 6/45 clean executions (13.3%), corresponding to 86.7% clean specificity. Five of the 15 unique conflict pairs were not serialized under the frozen dynamic declarations.

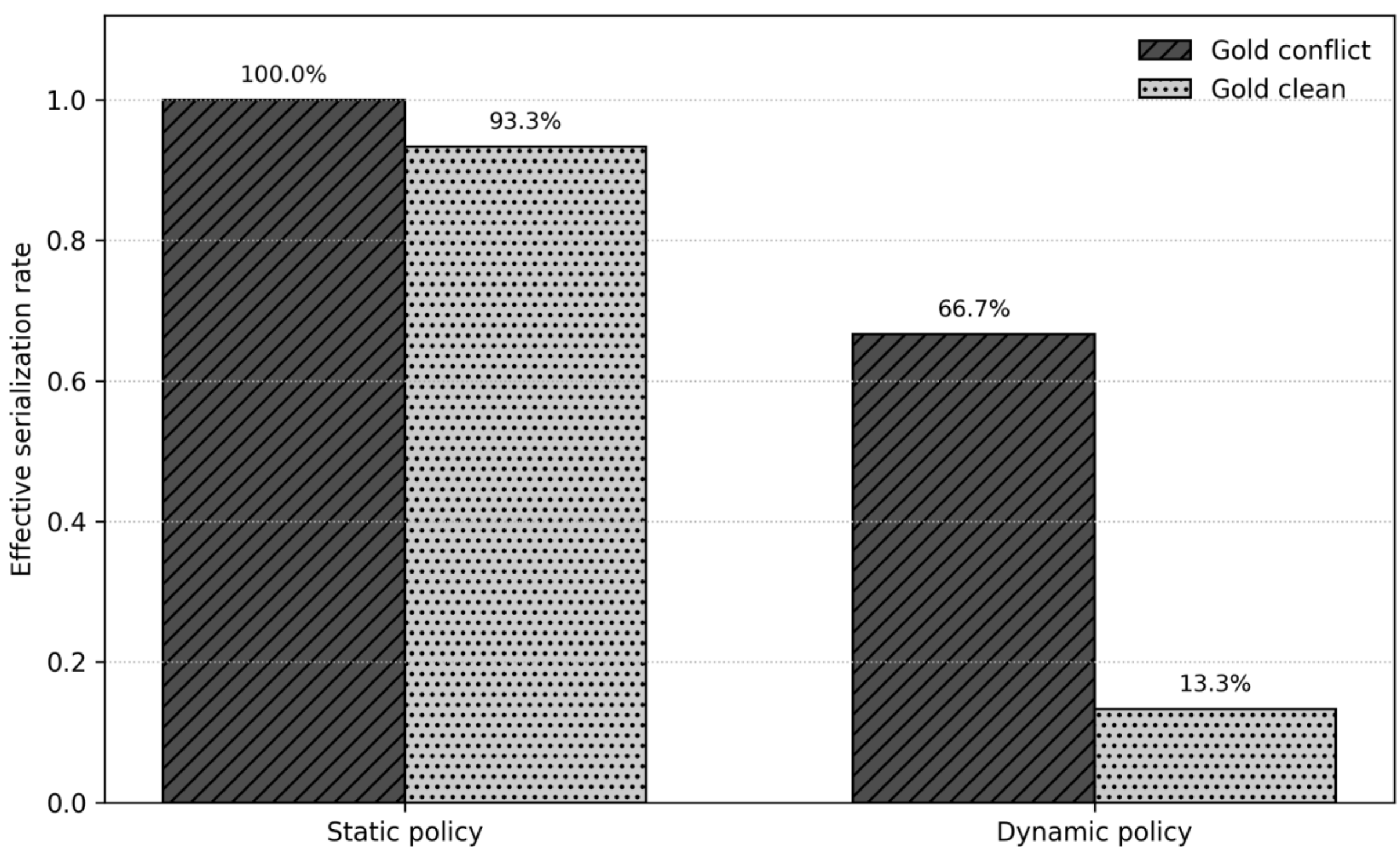

***Figure 3. Effective serialization by policy and benchmark label. Static admission was highly sensitive but poorly selective; dynamic admission preserved substantially more clean concurrency.***

## 4.4 Dynamic scope failure anatomy

| Mechanism | Static | Dynamic |
|---|---|---|
| Initial/effective serialization | 87 | 36 |
| Promotion attempts | 0 | 90 |
| Successful promotions | 0 | 38 |
| Rejected promotions | 0 | 0 |
| Undeclared-scope blocks | 0 | 46 |
| Scope-enforcement failures | 0 | 46 |
| Wasted dynamic steps | 0 | 317 |
| Wasted coder cost | $0.0000 | $0.3962 |

***Table 4. Claim Plane mechanism counts over 90 executions per policy.***

Dynamic scope failed closed in 46/90 executions and affected 20/30 unique feature pairs. Ten pairs failed under all three coder seeds, six under two seeds, and four under one seed. Twenty-five failures occurred in clean-labeled cases and 21 in conflict-labeled cases. Thirty occurred in executions that were initially parallel and 16 in executions that had already been serialized.

The failure shape is more specific than a generic planner miss. Forty-five of 46 blocked mutations targeted a file that already appeared somewhere in the corresponding frozen declaration; only one targeted an entirely undeclared file. Among the 45 failures with explicit line coordinates, 15 were within ten lines of a declared region and 30 were farther away. The blocks were concentrated in src/click/core.py (28), src/jinja2/ext.py (11), src/click/_termui_impl.py (4), and dirty_equals/_other.py (3). These data indicate that file-level recall was high but region-level coverage and runtime amendment were insufficient.

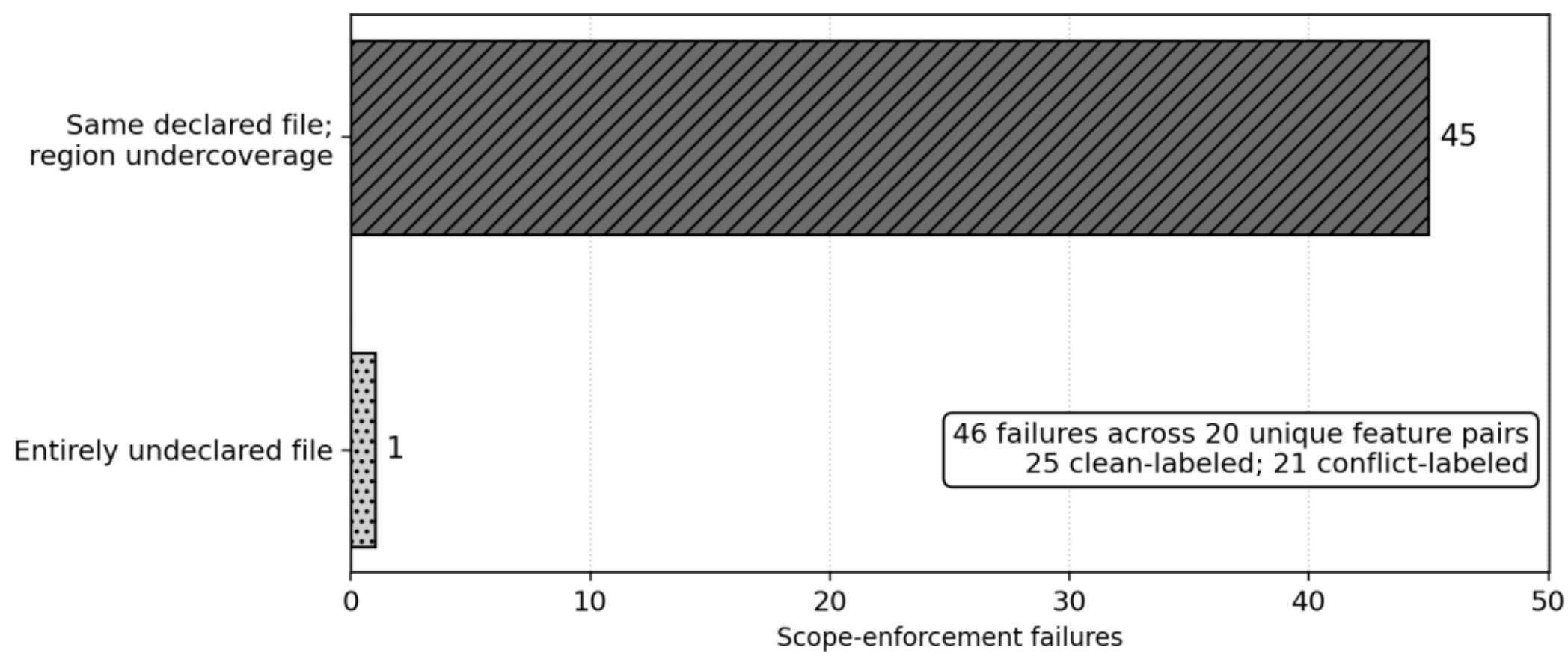

***Figure 4. Anatomy of dynamic scope-enforcement failures. Nearly every blocked mutation targeted a file already represented in the frozen declaration, but outside the admitted regions.***

No scope promotion was recorded as formally rejected. Instead, a mutation covered by a contingent region was promoted successfully, whereas a mutation that matched neither committed nor contingent authority failed immediately. This distinction matters for the next design: the system needs an explicit bounded amendment or replan state for legitimate newly discovered work, while retaining fail-closed behavior for genuinely unauthorized mutation.

## 4.5 Cost and logical critical path

| Arm | Coder total | Mean coder | Mean logical system | Logical critical path | Wasted dynamic cost |
|---|---|---|---|---|---|
| Parallel | $2.4888 | $0.0277 | $0.0277 | 134.2 s | $0.0000 |
| Static | $2.3189 | $0.0258 | $0.0923 | 188.0 s | $0.0000 |
| Dynamic | $1.6216 | $0.0180 | $0.0846 | 172.3 s | $0.3962 |
| Always serial | $2.2216 | $0.0247 | $0.0247 | 181.4 s | $0.0000 |

***Table 5. Logical model costs and critical-path estimates. These are study accounting measures, not a provider billing claim.***

The total logical coder cost was $8.6509. Including the planner cost once, total study logical cost was $10.6481. Dynamic coding cost was lower because many executions stopped early; this is not an efficiency advantage. The dynamic arm recorded $0.3962 of wasted coder cost before fail-closed termination.

Parallel execution had the shortest mean logical critical path at 134.2 seconds but the weakest reliability on conflict work. Static Claim Plane averaged 188.0 seconds, slightly above always serial at 181.4 seconds, because 87/90 executions were serialized and Claim Plane added planning overhead. Dynamic admission averaged 172.3 seconds, but its lower completion rate makes direct latency comparison misleading. The experiment therefore demonstrates a reliability trade-off, not a physical throughput gain.

## 4.6 Robustness across coder seeds

| Coder seed | Parallel | Static | Dynamic | Always serial |
|---|---|---|---|---|
| 101 | 7/30; 19/30 | 17/30; 29/30 | 6/30; 13/30 | 15/30; 29/30 |
| 202 | 6/30; 20/30 | 14/30; 30/30 | 7/30; 15/30 | 16/30; 28/30 |
| 303 | 8/30; 20/30 | 14/30; 28/30 | 7/30; 14/30 | 14/30; 29/30 |

***Table 6. Pair pass; integration success for each coder seed. Each cell reports successes out of 30.***

Static Claim Plane exceeded parallel pair pass under each seed: 17 versus 7 for seed 101, 14 versus 6 for seed 202, and 14 versus 8 for seed 303. Integration success was 29, 30, and 28 for static admission, compared with 19, 20, and 20 for parallel execution. Dynamic performance remained low across all three seeds, indicating that its failure was not driven by a single stochastic run.

## 5. Discussion

### 5.1 Pre-write admission recovered reliability

The strongest supported result is narrow but important: conservative claim-based admission recovered always-serial reliability on the evaluated conflict-labeled work. Static Claim Plane increased conflict pair pass by 53.3 percentage points and conflict integration by 62.2 points relative to unconstrained parallel execution. At every repository-task cluster, its pair-pass rate was at least that of parallel execution.

This does not mean that Claim Plane generated better code. Static and parallel outcomes were identical on the clean stratum. The control plane contributed by preventing two independently plausible trajectories from coexisting when their declared mutation authority overlapped.

### 5.2 Conservative admission is a reliability ceiling, not the final product

Static admission achieved its result by serializing 96.7% of executions. It therefore validates the safety end of the design space but does not satisfy the broader objective of serial-like quality with parallel throughput. A practical system must recover a meaningful subset of clean concurrency without reintroducing the conflict penalty.

### 5.3 Dynamic selectivity was promising, but the amendment path was not

Dynamic admission showed the desired classification signal: it separated conflict-labeled and clean-labeled work much more sharply than static admission. Yet selective initial admission was not enough. The coding agents frequently discovered mutations outside the frozen bounded regions, and the evaluated runtime had no safe continuation path beyond a fail-closed block.

The next deterministic engineering target is therefore Dynamic Scope v2: classify a newly observed mutation as a bounded local amendment, a newly discovered dependency, a contract expansion, a planner omission, or an unauthorized action; re-run admission; and continue, serialize, replan, escalate, or deny according to policy. The 46 failures in this study form a concrete regression corpus for that work.

### 5.4 Implications for a semantic dependency model

The results motivate, but do not yet validate, a learned semantic-dependency model (SDM). Static declarations had high conflict recall and almost no clean specificity. Dynamic declarations improved specificity but under-approximated the mutation surface. A compact SDM can sit between deterministic resource checks and a frontier semantic verifier, predicting operational outcomes such as safe parallelism, A-before-B, B-before-A, hard conflict, safe amendment, or uncertainty.

The current corpus is Dataset v0 rather than a sufficient training set. It contains only 30 unique semantic pairs, and the three coder seeds are correlated observations rather than independent examples. It is suitable for schema validation, deterministic or classical baselines, and data-collector design. A subsequent training corpus must use new repositories and must hold out entire tasks or repositories to avoid leakage.

### 5.5 Product implications

Claim Plane is useful without an SDM in two modes. First, a single-agent guard can enforce declared authority, explicit amendments, verification, and evidence. Second, a conservative swarm can fail closed and recover serial-level reliability where overlap is uncertain. Adaptive swarm execution should remain an experimental mode until the amendment path and selective-concurrency benchmark improve.

A production-facing evaluation should therefore compare a real coding runtime, such as Codex, with and without Claim Plane on previously unseen tasks. The key metric is not raw token price or agent count, but correct accepted deliveries per unit time and cost, together with undeclared mutations, human repair, and unnecessary serialization.

## 6. Related Work

### 6.1 Coordination benchmarks and orchestration

CooperBench [2] directly motivates the evaluated failure mode by showing a coordination penalty across more than 600 collaborative coding tasks. CAID [3] combines centralized dependency-aware delegation, isolated workspaces, and structured integration. Co-Coder [4] formalizes task decomposition as graph partitioning and reports accuracy, cost, and wall-clock gains when cross-agent dependencies are sparse. Claim Plane is complementary: it assumes candidate tasks or intents exist and governs whether their evolving mutation authority may coexist.

### 6.2 Runtime supervision, convergence, and repair

Shepherd [5] records reversible typed execution traces and uses meta-agent intervention to improve coding-pair success. CodeCRDT [6] coordinates through observable shared state and deterministic convergence. CoAgent [7] uses a pre-decided serialization order, notifications, and compensating repair by agents. These systems place control in runtime supervision, shared-state convergence, or optimistic repair. Claim Plane moves the primary decision boundary before governed mutation while retaining dependency invalidation and re-admission during execution.

### 6.3 Pre-write admission and deterministic control planes

ATM [8] is the closest contemporaneous pre-write system. It uses a CID broker, semantic atoms, bounded regions, fail-closed routing, and a neutral steward. Claim Plane does not claim priority over the general idea of pre-write admission. Its distinct formulation is versioned ChangeIntent authority with committed versus contingent scope, atomic promotion, intent-bound capability re-attestation, leases and fencing, and immutable patch evidence [1].

Madatha [9] also proposes a deterministic control plane, but for agent configuration supply chains, permissions, phase state, and prompt drift rather than concurrent repository mutation. SWE-agent [10] demonstrates that the agent-computer interface materially affects coding performance. Broader work argues for coordination as a separate architectural layer [11] and code as a verifiable agent harness [12]. The present study contributes empirical evidence for one specific authority-control mechanism within that larger systems agenda.

## 7. Threats to Validity

External validity is limited by 30 pairs, seven tasks, three Python project families, one planner model, one coder model, and oracle-localized initial context. Results may differ on other languages, larger repositories, longer tasks, or production runtimes. The benchmark conflict label is useful but does not exhaust semantic dependency types encountered in real software development.

Construct validity is limited by executable benchmark tests. Pair pass and integration success do not measure maintainability, security, architecture, or human review effort. The logical critical path is a modeled latency quantity, not an observed concurrent-provider speedup, because API calls were physically sequential.

Internal validity is strengthened by frozen planner output, fixed repository revisions, validated gold features, resumable shards, and hash-verified aggregation. However, the dynamic failures may reflect the interaction of this planner calibration, line-region coordinate mapping, coder behavior, and the specific fail-closed runtime rather than a fundamental limitation of contingent scope.

Statistical precision is constrained by seven repository-task clusters. Three coder seeds provide robustness to stochastic implementation behavior but do not create 90 independent semantic pairs. The clustered bootstrap respects this dependence, and the wide intervals should remain visible in interpretation.

Finally, Claim Plane can only enforce mutations that cross its observable boundary. A complete broker or OS-level monitor supports stronger authority claims than post-hoc diff verification. Adapter integrations must state their actual interception and bypass guarantees.

## 8. Conclusion

This confirmatory study shows that deterministic claim-based pre-write admission can protect parallel coding work from integration interference. Static Claim Plane more than doubled pair pass relative to unconstrained parallel execution, restored conflict-labeled integration to the always-serial level, and matched always serial on the primary outcome at every task cluster.

The same result establishes the limitation of the current policy: it serialized nearly every execution, including clean work. Dynamic admission recovered much better selectivity but failed operationally because undeclared region-level mutations were blocked in more than half of executions. Useful parallelism therefore requires two additional capabilities: robust bounded scope amendment and more accurate semantic dependency prediction.

The Claim Plane thesis is consequently narrower and stronger after this study. Pre-write admission is a viable reliability mechanism. Serial-like reliability with parallel throughput remains an open systems objective. The released corpus provides a reproducible basis for Dynamic Scope v2, an SDM data pipeline, and an external Codex evaluation on previously unseen software tasks.

## Data and Code Availability

The public Claim Plane source repository is available at github.com/SkeinRank/claim-plane. The complete public reproducibility dataset is available at huggingface.co/datasets/skeinrank/claim-plane-confirmatory-30x3.

The released archive is claim-plane-confirmatory-30x3-public.tar.gz. Its SHA-256 digest is 9a8c7f7593d1569577be319e79bbf9af8d890b582343c99f28dd4c00626f89bf. The internal study fingerprint is df01b32b75261331b922a1276d82c66dfb4e1d0134c0b2b6d32c6f3c37b5d1c7. The publication manifest records 360 executions, nine run IDs, 5,000 bootstrap samples, and hashes for all analysis inputs and outputs. The companion architecture paper is arXiv:2607.21909, DOI 10.48550/arXiv.2607.21909.

## Generative-AI Tool-Use Statement

Generative AI tools were used as assistive tools during software development, data inspection, and manuscript preparation. The author reviewed the final manuscript, and all reported experimental results were derived from and checked against the released study artifacts.

## Appendix A. Reproducibility Record

| Field | Value |
|---|---|
| Study ID | claim-plane-confirmatory-30x3 |
| Study fingerprint | df01b32b75261331b922a1276d82c66dfb4e1d0134c0b2b6d32c6f3c37b5d1c7 |
| Claim Plane study metadata version | 0.2.1 |
| Claim Plane execution commit | fb32a36f743c658b1636fca2f47b3c9bc07a4cc4 |
| Execution image used by launcher | claim-plane-cooperbench:0.9.4 |
| Container environment | cooperbench-linux-v1 |
| Architecture | aarch64 |
| Python runtime | 3.12.13 |
| CooperBench commit | d46d9e73fa64159e0428b480f293623de90be1ad |
| Frozen dataset SHA-256 | c5a7638846b4079a4e31cc2c40d7fad2728194c8b165c8f553211f39a7850cac |
| Pair-selection seed | 42 |
| Planner model / seed | deepseek/deepseek-v4-pro / 1701 |
| Coder model / seeds | deepseek/deepseek-v4-flash / 101, 202, 303 |
| Pairs / arms / executions | 30 / 4 / 360 |
| Shards | 3 per seed; 9 total |
| Bootstrap | 5,000 samples; seed 20260727; cluster = repo + task_id |
| One-time planner logical cost | $1.99 |
| Total study logical cost | $10.64 |
| Dataset repository | huggingface.co/datasets/skeinrank/claim-plane-confirmatory-30x3 |
| Public archive | claim-plane-confirmatory-30x3-public.tar.gz |
| Public archive SHA-256 | 9a8c7f7593d1569577be319e79bbf9af8d890b582343c99f28dd4c00626f89bf |

***Table A1. Fixed identifiers and reproducibility parameters for the confirmatory study.***

The publication verifier reported 15 derived files and 20 input files verified with zero mismatches. All nine shards were complete before aggregation. Frozen plans were shared across the three coder seeds and the static and dynamic Claim Plane arms.